\documentclass[twocolumn]{svjour3}
\usepackage{graphics}
\usepackage{amsmath,amssymb}

\journalname{Eur. Phys. J. A}

\begin{document}

\title{$ND$ and $NB$ systems in quark delocalization color screening model}

\author{Lifang Zhao \and Hongxia Huang \and Jialun Ping\thanks{Corresponding
 author: jlping@njnu.edu.cn}}

\institute{Lifang Zhao \at Department of Quality-Oriented Education, Nanjing College of Information Technology,
Nanjing 210023, P.R. China \and Hongxia Huang \and Jialun Ping \at Department of Physics, Nanjing Normal University,
Nanjing 210023, P. R. China}

\titlerunning{$ND$ and $NB$ systems in QDCSM}
\authorrunning{Zhao, Huang,Ping}

\maketitle

\begin{abstract}
The $ND$ and $NB$ systems with $I=0$ and $1$, $J^{P}=\frac{1}{2}^{\pm}$, $\frac{3}{2}^{\pm}$,
and $\frac{5}{2}^{\pm}$ are investigated within the framework of quark delocalization color
screening model. The results show that all the positive parity states are unbound.
By coupling to the $ND^{*}$ channel, the state $ND$ with $I=0,~J^{P}=\frac{1}{2}^{-}$ can form
a bound state, which can be invoked to explain the observed $\Sigma(2800)$ state. The mass of
the $ND^{*}$ with $I=0,~J^{P}=\frac{3}{2}^{-}$ is close to that of the reported $\Lambda_{c}(2940)^{+}$,
which indicates that $\Lambda_{c}(2940)^{+}$ can be explained as a $ND^{*}$ molecular state
in QDCSM. Besides, the $\Delta D^{*}$ with $I=1,~J^{P}=\frac{5}{2}^{-}$ is also a possible
resonance state. The results of the bottom case of $NB$ system are similar to those of the
$ND$ system. Searching for these states will be a challenging subject of experiments.
\end{abstract}

\section{\label{sec:introduzione}Introduction}

In the past decade, many near-threshold charmonium-like states have been reported by Belle,
BaBar, BESIII, LHCb and other collaborations, which triggers lots of studies on the
molecule-like hadrons containing heavy quarks. For example, the triplet of excited $\Sigma_{c}$
baryons, $\Sigma_{c}(2800)$, was observed by Belle~\cite{Belle1}, and they tentatively identified
the quantum numbers of these states as $J^{P}=\frac{3}{2}^{-}$.
The same neutral state $\Sigma_{c}^{0}$ was also observed in $B$ decays by the BaBar
collaboration with the mean value of mass higher by about $3\sigma$ from that obtained
by Belle~\cite{BABAR1}, although the widths from these two measurements are consistent.
Moreover, a new charmed hadron $\Lambda_{c}(2940)^{+}$ with mass $M = 2939.8 \pm
1.3(\mbox{stat}) \pm 1.0(\mbox{syst})~\mbox{MeV/c}^2$ and width $\Gamma = 17.5 \pm 5.2(\mbox{stat})
\pm 5.9(\mbox{syst})~\mbox{MeV/c}^2$ was reported by BaBar collaboration by analyzing the
$D^{0}p$ invariant mass spectrum~\cite{BABAR2}, and it is confirmed as a resonant structure
in the final state of $\Sigma_{c}(2455)^{0,++}\pi^{\pm} \rightarrow \Lambda_{c}^{+}\pi^{+}\pi^{-}$
by Belle~\cite{Belle2}.

The experimental observations have stimulated extensive interest in understanding the
structures of the states $\Sigma_{c}(2800)$ and $\Lambda_{c}(2940)^{+}$. Since the
$\Sigma_{c}(2800)$ and $\Lambda_{c}(2940)^{+}$ are near the threshold of $ND$ and $ND^{*}$,
respectively, many work treat them as candidates of molecular states. For the
$\Sigma_{c}(2800)$, M. Lutz and E. Kolomeitsev interpreted it as a chiral molecule~\cite{Lutz},
while C. Jim\'{e}nez-Tejero {\em et al.} found it was a dynamically generated
resonance with a dominant $ND$ configuration~\cite{Jim}. Y. B. Dong {\em et al.} estimated
the strong $\Lambda_{c}\pi$ decays of the $\Sigma_{c}(2800)$ state for different spin-parity
assignments by assuming a dominant molecular $ND$ structure of the state and showed that
the decay widths of $\Sigma_{c} \rightarrow \Lambda_{c}\pi$ were consistent with current data
for the $J^{P} = \frac{1}{2}^{+}$ and $J^{P} = \frac{3}{2}^{-}$ assignments~\cite{Dong}.
Moreover, J. R. Zhang investigated $\Sigma_{c}(2800)$ and $\Lambda_{c}(2940)^{+}$ as the
$S-$wave $ND$ state with $J^{P} = \frac{1}{2}^{-}$ and $ND^{*}$ state with $J^{P} = \frac{3}{2}^{-}$,
respectively in the framework of QCD sum rules. The results showed that the masses of these
two states were bigger than the experimental data, but the compact structure of the states
could be ruled out~\cite{Zhang}. For the $\Lambda_{c}(2940)^{+}$, X. G. He {\em et~al.}
proposed that it may be a $D^{*0}p$ molecular state with $J^{P} = \frac{1}{2}^{-}$~\cite{HeXG},
while Y. B. Dong {\em et~al.} showed the $\Lambda_{c}(2940)^{+}$ was a molecular state composed
of a nucleon and $D^{*}$ mesons with $J^{P} = \frac{1}{2}^{+}$ by studying the decay widths
of the strong two-body decay channels $\Lambda_{c}(2940)^{+} \rightarrow p D^{0}$,
$\Sigma_{c}^{++}\pi^{-}$ and $\Sigma_{c}^{0}\pi^{+}$~\cite{Dong1}, and this conclusion was
confirmed by investigating the width of the radiative decay process $\Lambda_{c}(2940)^{+}
\rightarrow \Lambda_{c}(2286)^{+}+\gamma$~\cite{Dong2} and the strong three-body decay process
$\Lambda_{c}(2940)^{+} \rightarrow \Lambda_{c}(2286)^{+}\pi^{+}\pi^{-}$ and
$\Lambda_{c}(2286)^{+}\pi^{0}\pi^{0}$~\cite{Dong3}. Moreover, J. He and X. Liu explained
the $\Lambda_{c}(2940)^{+}$ as an isoscalar $S-$wave or $P-$wave $D^{*}N$ system with $J^{P}
= \frac{3}{2}^{-}$ or $J^{P} = \frac{1}{2}^{+}$ in the framework of the one-boson-exchange
model~\cite{HeJ}. And in Ref.~\cite{Gar}, they found a possible molecular candidate with
$J^{P} = \frac{3}{2}^{-}$ for the $\Lambda_{c}(2940)^{+}$. In addition, a bound state $D^{*}N$
with $J^{P} = \frac{3}{2}^{-}$, which can be explained as the $\Lambda_{c}(2940)^{+}$, was also
obtained in a constituent quark model~\cite{Ortega}. In the work of Ref.~\cite{XieJJ},
the total cross section of the $\pi^{-}p \rightarrow D^{-}D^{0}p$ reaction was calculated within
an effective lagrangian approach, which indicated that the spin-parity assignment of
$\frac{1}{2}^{-}$ for $\Lambda_{c}(2940)^{+}$ gave a sizable enhancement for the total cross section
in comparison with a choice of $J^{P} =\frac{1}{2}^{+}$.

Another way to describe the states $\Sigma_{c}(2800)$ and $\Lambda_{c}(2940)^{+}$ is based on
the assumption that they are conventional charmed baryons. A relativized potential model predicted
that the masses of $\Sigma_{c}^{*}$ with $J^{P} = \frac{3}{2}^{-}$ or $\frac{5}{2}^{-}$ and
$\Lambda_{c}^{*}$ with $J^{P} = \frac{3}{2}^{+}$ or $\frac{5}{2}^{-}$ are close to the value of
$\Sigma_{c}(2800)$ and $\Lambda_{c}(2940)^{+}$, respectively~\cite{Capstick}. The strong decays of
$\Sigma_{c}(2800)$ and $\Lambda_{c}(2940)^{+}$ as charmed baryons have been studied by using
$^{3}P_{0}$ model~\cite{ChenC}, heavy hadron chiral perturbation theory~\cite{ChengHY}, and
chiral quark model~\cite{ZhongXH}. Moreover, Ebert {\em et~al.} suggested $\Sigma_{c}(2800)$ as
one of the orbital ($1P$) excitations of the $\Sigma_{c}$ with $J^{P} = \frac{1}{2}^{-}$,
$\frac{3}{2}^{-}$ or $\frac{5}{2}^{-}$, and proposed $\Lambda_{c}(2940)^{+}$ as the first radial
excitation of $\Sigma_{c}$ with $J^{P} = \frac{3}{2}^{+}$~\cite{Ebert}. J. He {\em et~al.} estimated
the production rate of $\Lambda_{c}(2940)^{+}$ as a charmed baryon at PANDA~\cite{HeJ2}.
H. Garcilazo {\em et~al.} solved exactly the three-quark problem by means of the Faddeev method
in momentum space, and showed that $\Sigma_{c}(2800)$ would correspond to an orbital excitation
with $J^{P} = \frac{1}{2}^{-}$ or $\frac{3}{2}^{-}$, and the $\Lambda_{c}(2940)^{+}$ may
constitute the second orbital excitation of the $\Lambda_{c}$ baryon~\cite{Garcilazo}.

Although many theoretical explanations to $\Sigma_{c}(2800)$ and $\Lambda_{c}(2940)^{+}$
were proposed, the properties of these two states are still in ambiguous. Therefore, more
efforts are needed to reveal the underlying structure of these two states. Quantum chromodynamics
(QCD) is widely accepted as the fundamental theory of the strong interaction. However, the direct
use of QCD for low-energy hadron physics, for example, the properties of hadrons, the
nucleon-nucleon ($NN$) interaction, is still difficult because of the nonperturbative complications
of QCD. QCD-inspired quark models are still the main approach to study the hadron-hadron interaction.

It is well known that the forces between nucleons (hadronic clusters of quarks) are qualitative
similar to the forces between atoms (molecular force). This similarity is naturally explained in
the quark delocalization color screening model (QDCSM)~\cite{QDCSM0}, which has been developed and
extensively studied. In this model, quarks confined in one nucleon are allowed to delocalize to a
nearby nucleon and the confinement interaction between quarks in different baryon orbits is
modified to include a color screening factor. The latter is a model description of the hidden color
channel coupling effect~\cite{QDCSM1}. The delocalization parameter is determined by the dynamics
of the interacting quark system, thus allows the quark system to choose the most favorable
configuration through its own dynamics in a larger Hilbert space. The model gives a good
description of nucleon-nucleon and hyperon-nucleon interactions and the properties of
deuteron~\cite{QDCSM2}. It is also employed to calculated the baryon-baryon scattering phase
shifts and the dibaryon candidates in the framework of the resonating group method
(RGM)~\cite{QDCSM3,QDCSM4}. Recently, it has been used to investigate the pentaquarks with
heavy quarks, and the $P_{c}(4380)$ can be explained as the molecular pentaquark of
$\Sigma^{*}_{c}D$ with quantum numbers $IJ^{P}=\frac{1}{2}\frac{3}{2}^{-}$ in QDCSM~\cite{HuangHX}.

In present work, QDCSM is employed to study the properties of $ND$ systems, and the
channel-coupling effect of $ND^{*}$, $\Delta D$, and $\Delta D^{*}$ channels are included.
Our purpose is to investigate whether $\Sigma_{c}(2800)$ and $\Lambda_{c}(2940)^{+}$ could
be explained as a molecular state composed of a nucleon and $D$ or $D^{*}$ mesons.
On the other hand, we also want to see if any other bound or resonance state exist or not.
Extension of the study to the bottom case is also interesting and is performed here.
The structure of this paper is as follows. After the introduction, we present a brief
introduction of the quark model used in section 2. Section 3 devotes to the numerical
results and discussions. The summary is shown in the last section.

\section{The quark delocalization color screening
model (QDCSM)}
The detail of QDCSM used in the present work can be found  in the
references~\cite{QDCSM0,QDCSM1,QDCSM2,QDCSM3,QDCSM4}. Here, we
just present the salient features of the model. The model
Hamiltonian is:
\begin{eqnarray}
H &=& \sum_{i=1}^6 \left(m_i+\frac{p_i^2}{2m_i}\right) -T_c
+\sum_{i<j} V_{ij}, \\
V_{ij} & = &  V^{G}(r_{ij})+V^{\chi}(r_{ij})+V^{C}(r_{ij}),  \nonumber \\
V^{G}(r_{ij})&=& \frac{\alpha_{s}}{4} \boldsymbol{\lambda}_i \cdot
\boldsymbol{\lambda}_j
\left[ \frac{1}{r_{ij}}-\frac{\pi}{2}\delta(\boldsymbol{r}_{ij})
   \left(\frac{1}{m_{i}^{2}} +\frac{1}{m_{j}^{2}} \right. \right. \nonumber \\
& & \left. \left.~~~~~~~~~~~~~~+\frac{4\boldsymbol{\sigma}_i\cdot
 \boldsymbol{\sigma}_j}{3m_{i}m_{j}}  \right)-\frac{3}{4m_{i}m_{j}r^3_{ij}}S_{ij}\right],
\nonumber \\
V^{\chi}(r_{ij})&=& \frac{\alpha_{ch}}{3}
\frac{\Lambda^2 m_\chi}{\Lambda^2-m_{\chi}^2} \left\{ \left[
Y(m_\chi r_{ij})- \frac{\Lambda^3}{m_{\chi}^3}Y(\Lambda r_{ij})
\right] \right. \nonumber \\
&&  \boldsymbol{\sigma}_i \cdot\boldsymbol{\sigma}_j
\left. +\left[ H(m_\chi r_{ij})-\frac{\Lambda^3}{m_\chi^3}
H(\Lambda r_{ij})\right] S_{ij} \right\} \nonumber \\
& & {\mathbf F}_i \cdot
{\mathbf F}_j, ~~~\chi=\pi,K,\eta \nonumber
\end{eqnarray}
\begin{eqnarray}
V^{C}(r_{ij})&=& -a_c {\mathbf \lambda}_i \cdot {\mathbf
\lambda}_j [f(r_{ij})+V_0], \nonumber
\\
 f(r_{ij}) & = &  \left\{ \begin{array}{ll}
 r_{ij}^2 &
 \qquad \mbox{if }i,j\mbox{ occur in the same} \\
  & \mbox{~~~~~~baryon orbit} \\
  \frac{1 - e^{-\mu_{ij} r_{ij}^2} }{\mu_{ij}} & \qquad
 \mbox{if }i,j\mbox{ occur in different} \\
   & \mbox{~~~~~~baryon orbits} \\
 \end{array} \right.
\nonumber \\
S_{ij} & = &  \frac{{\mathbf (\sigma}_i \cdot {\mathbf r}_{ij})
({\mathbf \sigma}_j \cdot {\mathbf
r}_{ij})}{r_{ij}^2}-\frac{1}{3}~{\mathbf \sigma}_i \cdot {\mathbf
\sigma}_j. \nonumber
\end{eqnarray}
Where $S_{ij}$ is quark tensor operator; $Y(x)$ and $H(x)$ are
standard Yukawa functions~\cite{Valcarce}; $T_c$ is the kinetic
energy of the center of mass; $\alpha_{ch}$ is the chiral coupling
constant, determined as usual from the $\pi$-nucleon coupling
constant; $\alpha_{s}$ is the quark-gluon coupling constant. In
order to cover the wide energy range from light to heavy
quarks one introduces an effective scale-dependent quark-gluon
coupling $\alpha_{s}(\mu)$\cite{Vijande}:
\begin{equation}
\alpha_{s}(\mu) =
\frac{\alpha_{0}}{\ln(\frac{\mu^2+\mu_{0}^2}{\Lambda_{0}^2})},
\end{equation}
where $\mu$ is the reduced mass of two interacting quarks. All other symbols have their usual meanings.
Here, a phenomenological color screening confinement potential is used, and $\mu_{ij}$ is the color
screening parameter. For the light-flavor quark system, it is determined by fitting the deuteron
properties, $NN$ scattering phase shifts, $N\Lambda$ and $N\Sigma$ scattering phase shifts,
respectively, with $\mu_{uu}=0.45$, $\mu_{us}=0.19$ and $\mu_{ss}=0.08$, satisfying the relation,
$\mu_{us}^{2}=\mu_{uu}*\mu_{ss}$. When extending to the heavy quark case, there is no experimental data available, so
we take it as a common parameter. In the present work, we take
$\mu_{cc}=0.01~{\rm fm}^{-2}$ and $\mu_{uc}=0.067~{\rm
fm}^{-2}$, also satisfy the relation
$\mu^{2}_{uc}=\mu_{uu}*\mu_{cc}$. All other parameters are also taken from our previous
work~\cite{QDCSM4}, except for the charm and bottom quark masses $m_{c}$ and $m_{b}$, which are
fixed by a fitting to the masses of the charmed and bottom mesons. The values of
those parameters are listed in Table~\ref{parameters}. The corresponding masses of the baryons and charmed
and bottom mesons are shown in Table~\ref{baryon}.

\begin{table}[h]
\begin{center}
\caption{Model parameters:
$m_{\pi}=0.7~{\rm fm}^{-1}$, $m_{k}=2.51~{\rm fm}^{-1}$,
$m_{\eta}=2.77~{\rm fm}^{-1}$, $\Lambda_{\pi}=4.2~{\rm fm}^{-1}$,
$\Lambda_{k}=5.2~{\rm fm}^{-1}$, $\Lambda_{\eta}=5.2~{\rm
fm}^{-1}$, $\alpha_{ch}=0.027$.}
{\begin{tabular}{cccccc} \hline
$b$ & $m_{s}$ & $m_{c}$ & $m_{b}$ & $ a_c$  \\
$({\rm fm})$ & $({\rm MeV})$ & $({\rm MeV})$ & $({\rm MeV})$ & $({\rm MeV\,fm}^{-2})$  \\ \hline
0.518  &  573 & 1675 &    5086  &  58.03 \\ \hline
$V_{0}$ &  ~$\alpha_{0}$~ &  ~$\Lambda_{0}$~ & ~$\mu_{0}$~ &  \\
(MeV) &  ~~ &  ~$({\rm fm}^{-1})$~ & ~$({\rm MeV})$~ &  \\ \hline
 -1.2883 &  0.5101 &  1.525 &   445.808  & \\
\hline
\end{tabular}}
\label{parameters}

\caption{\label{mass}The masses of the baryons and charmed
and bottom mesons (in MeV).}
{\begin{tabular}{ccccccc}\hline
        & $N$     & $\Delta$ & $\Lambda$ & $\Sigma$  & $\Sigma^*$  & $\Xi$    \\ \hline
 QDCSM  & 939     & 1232     & 1118      & 1224      & 1358        & 1365     \\
 Exp.   & 939     & 1232     & 1116      & 1193      & 1385        & 1318     \\ \hline
        & $\Xi^*$ & $\Omega$ & $D$       & $D^{*}$   &  $B$        & $B^{*}$  \\ \hline
 QDCSM  & 1499    & 1654     &  1865     & 1900      & 5279        & 5290    \\
 Exp.   & 1533    & 1672     & 1864      & 2007      & 5279        & 5325 \\
 \hline
\end{tabular}}
\label{baryon}
\end{center}
\end{table}

The quark delocalization in QDCSM is realized by specifying the single particle orbital
wave function of QDCSM as a linear combination of left and right Gaussians, the single
particle orbital wave functions used in the ordinary quark cluster model,
\begin{eqnarray}
\psi_{\alpha}(\mathbf{s}_i ,\epsilon) & = & \left(
\phi_{\alpha}(\mathbf{s}_i)
+ \epsilon \phi_{\alpha}(-\mathbf{s}_i)\right) /N(\epsilon), \nonumber \\
\psi_{\beta}(-\mathbf{s}_i ,\epsilon) & = &
\left(\phi_{\beta}(-\mathbf{s}_i)
+ \epsilon \phi_{\beta}(\mathbf{s}_i)\right) /N(\epsilon), \nonumber \\
N(\epsilon) & = & \sqrt{1+\epsilon^2+2\epsilon e^{-s_i^2/4b^2}}. \label{1q}
\end{eqnarray}
\begin{eqnarray}
\phi_{\alpha}(\mathbf{s}_i) & = & \left( \frac{1}{\pi b^2}
\right)^{3/4}
   e^{-\frac{1}{2b^2} (\mathbf{r}_{\alpha} - \mathbf{s}_i/2)^2} \nonumber \\
\phi_{\beta}(-\mathbf{s}_i) & = & \left( \frac{1}{\pi b^2}
\right)^{3/4}
   e^{-\frac{1}{2b^2} (\mathbf{r}_{\beta} + \mathbf{s}_i/2)^2}. \nonumber
\end{eqnarray}
Here $\mathbf{s}_i$, $i=1,2,...,n$ are the generating coordinates, which are
introduced to expand the relative motion wavefunction~\cite{QDCSM1}. The mixing parameter
$\epsilon(\mathbf{s}_i)$ is not an adjusted one but determined variationally by the dynamics
of the multi-quark system itself. This assumption allows the multi-quark system to choose
its favorable configuration in the interacting process. It has been used to explain the
cross-over transition between hadron phase and quark-gluon plasma phase~\cite{Xu}.

\section{The Results and Discussions}
Here, we investigate the $ND$ systems with $I=0$ and $1$, $J^{P}=\frac{1}{2}^{\pm}$,
$\frac{3}{2}^{\pm}$, and $\frac{5}{2}^{\pm}$. For the negative parity states, the orbital
angular momentum $L$ between clusters is set to 0; and for the positive parity states,
$L=1$. All the channels involved are listed in Table~\ref{channels}. The channel coupling
calculation is also performed. However, we find there is no any bound state with the
positive parity in our calculations. In the following we only show the results of
the negative parity states.
\begin{table}[h]
\begin{center}
\caption{The channels involved in the calculation.}
{\begin{tabular}{@{}cccccccccccccc@{}} \hline
 ~~~$I=0,~S=1/2$~~~ & $ND$, & $ND^{*}$   \\
 ~~~$I=0,~S=3/2$~~~ & $ND^{*}$   \\
 ~~~$I=1,~S=1/2$~~~ & $ND$, & $ND^{*}$,  & $\Delta D^{*}$   \\
 ~~~$I=1,~S=3/2$~~~ & $ND^{*}$, & $\Delta D$, & $\Delta D^{*}$   \\
 ~~~$I=1,~S=5/2$~~~ & $\Delta D^{*}$   \\
 \hline
\end{tabular}}
\label{channels}
\end{center}
\end{table}

Because an attractive potential is necessary for forming bound state or resonance,
we first calculate the effective potentials of all the channels listed in Table~\ref{channels}.
The effective potential between two colorless clusters is defined as, $V(s)=E(s)-E(\infty)$,
where $E(s)$ is the energy of the system at the separation $s$ of two clusters, which is
obtained by the adiabatic approximation. The effective potentials of the $S$-wave $ND$ systems
with $I=0$ and $I=1$ are shown in Fig. 1 and 2, respectively.
From Fig. 1(a), we can see that the potential of the $J^{P}=\frac{1}{2}^{-}$ channel $ND$ is
weak attractive, while for the channel $ND^{*}$, the potential is repulsive and so no bound
state can be formed in these two single channels. However, the attractions of
$J^{P}=\frac{3}{2}^{-}$ $ND^{*}$ is much larger as shown in Fig. 1(b), which means that
two hadrons, $N$ and $D^{*}$, could be bound together in this case.
For the effective potentials of the $I=1$ system as shown in Fig. 2, the attractions are
large for all $\Delta D^{*}$ channels, as well as the $J^{P}=\frac{3}{2}^{-}$ $\Delta D$
channel, followed by the $J^{P}=\frac{1}{2}^{-}$ $ND^{*}$ channel, the potential of which
is very weak, while for both $J^{P}=\frac{1}{2}^{-}$ $ND$ and $J^{P}=\frac{3}{2}^{-}$ $ND^{*}$
channels, the potentials are repulsive.

\begin{figure}
\begin{center}
\resizebox{0.48\textwidth}{!}{\includegraphics{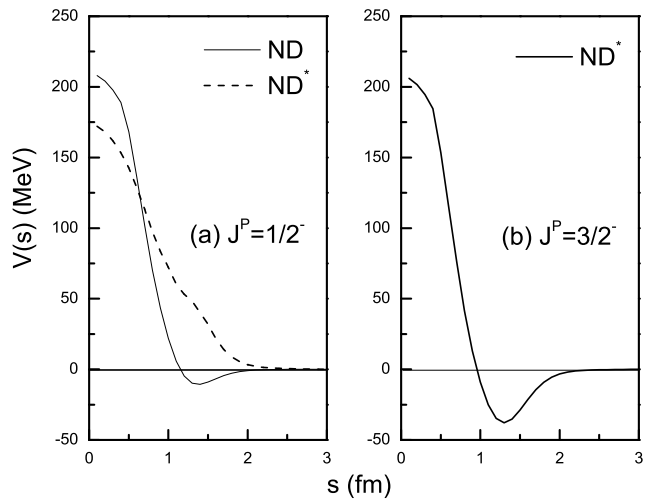}}

\caption{The potentials of different channels for the
$ND$ system with $I=0$.}

\end{center}
\end{figure}

\begin{figure*}
\begin{center}
\resizebox{0.9\textwidth}{!}{\includegraphics{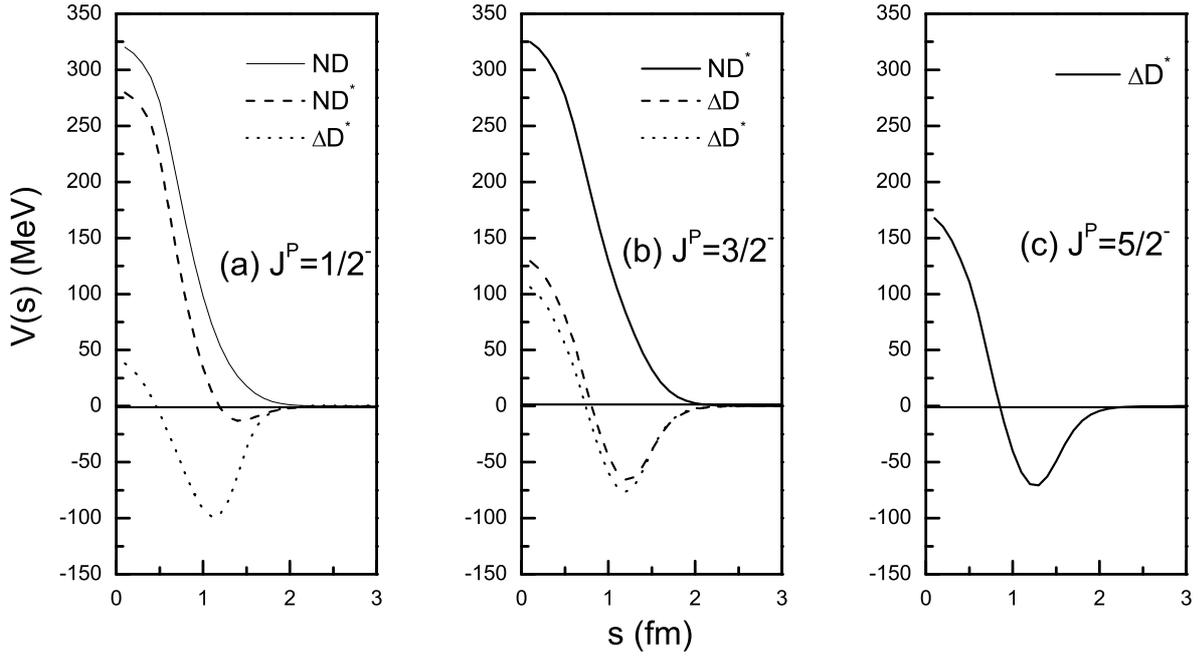}}

\caption{The potentials of different channels for the
$ND$ system with $I=1$.}

\end{center}
\end{figure*}

In order to check whether the possible bound states can be realized, a dynamic
calculation is needed. Here the RGM equation is employed. Expanding the relative motion
wavefunction between two clusters in the RGM equation by gaussians, the integro-differential
equation of RGM can be reduced to an algebraic equation, the generalized eigen-equation.
The energy of the system can be obtained by solving the eigen-equation. In the calculation,
the baryon-meson separation ($|\mathbf{s}_n|$) is taken to be less than 6 fm (to keep the
matrix dimension manageably small). The binding energies and the masses of every single
channel and those with channel coupling are listed in Table~\ref{bound_c}.
\begin{table*}[ht]
\begin{center}
\caption{The binding energies ($E_B$) and the masses ($M$) (in MeV) of every single channels and those of channel
coupling ($c.c.$) for the $ND$ system.}
\begin{tabular}{lccccccc}
\hline 
  & & ~~~$ND$~~  & ~~$ND^{*}$~~ & ~~$\Delta D$~~ & ~~$\Delta D^{*}$~~ & ~$c.c.$~\\
\hline
 $IJ^{P}=0\frac{1}{2}^{-}$ & ~~$E_B/M$~~ & $ub/2803$ & $ub/2946$  & $-/-$ & $-/-$ & $-2.0/2801.0$  \\ \noalign{\smallskip}
 $IJ^{P}=0\frac{3}{2}^{-}$ & ~~$E_B/M$~~ &$-/-$ & $-5.7/2940.3$  & $-/-$ & $-/-$ & $-5.7/2940.3$   \\ \noalign{\smallskip}
 $IJ^{P}=1\frac{1}{2}^{-}$ & ~~$E_B/M$~~ &$ub/2803$ & $ub/2946$  & $-/-$ & $-25.3/3213.7$ & $ub/2803$ \\ \noalign{\smallskip}
 $IJ^{P}=1\frac{3}{2}^{-}$ & ~~$E_B/M$~~ &$-/-$ & $ub/2946$  & $-14.7/3081.3$ & $-18.5/3220.5$ & $ub/2946$    \\ \noalign{\smallskip}
 $IJ^{P}=1\frac{5}{2}^{-}$ & ~~$E_B/M$~~ &$-/-$ & $-/-$  & $-/-$ & $-28.9/3210.1$ & $-28.9/3210.1$    \\
\hline 
\end{tabular}
\label{bound_c}
\end{center}
\end{table*}

For the $I=0,~J^{P}=\frac{1}{2}^{-}$ system, the single channel calculation shows that
the energy of $ND$ is above its threshold although there is a weak attraction between $N$ and $D$.
It is unbound (labeled as "$ub$" in Table~\ref{bound_c}), and the $ND^{*}$ is also unbound,
because the interaction between $N$ and $D^{*}$ is repulsive as mentioned above. However,
by taking into account the channel-coupling effect, we obtain a stable state, the mass of which
is lower than the threshold of $ND$. The binding energy and the mass of this bound state is shown
in Table~\ref{bound_c}, which is labeled as "$c.c.$".
First, we should mention how we obtain the mass of a bound molecular pentaquark. Generally,
the mass of a molecular pentaquark can be written as $M^{the.}=M^{the.}_{1}+M^{the.}_{2}+B$,
where $M^{the.}_{1}$ and $M^{the.}_{2}$ stand for the theoretical masses of a baryon and
a meson respectively, and $B$ is the binding energy of this molecular state. In order to
minimize the theoretical errors and to compare calculated results to the experimental data,
we shift the mass of molecular pentaquark to $M=M^{exp.}_{1}+M^{exp.}_{2}+B$, where the
experimental masses of baryons and mesons are used. Taking this bound state as an example,
the calculated mass of pentaquark is $2801.6$ MeV, then the binding energy $B$ is obtained by
subtracting the theoretical masses of $N$ and $D$, $2801.6-939.0-1864.6=-2.0$ (MeV).
Adding the experimental masses of the hadrons, the mass of the pentaquark
$M=939.0+1864.0+(-2.0)=2801.0$ (MeV) is arrived.
Secondly, we find that the mass of this bound state is close to the mass of the observed
$\Sigma(2800)$, which was reported by Belle collaboration. Therefore, in our quark model
calculation $\Sigma(2800)$ can be explained as a $ND$ molecular state with the quantum number
$J^{P}=\frac{1}{2}^{-}$. Finally, the coupling between the $S-$wave $ND$ and $ND^{*}$ channels,
which is through the central force, is of crucial importance for obtaining a bound state here.
In order to see the strength of these channel-coupling, we calculate the transition potential
of these two channels, which is shown in Fig. 3(a). Obviously, it is a strong coupling among
these channels that makes $ND$ the bound state. This mechanism to form a bound state has been
proposed before. Eric S. Swanson proposed that the admixtures of $\rho J/\psi$ and
$\omega J/\psi$ states were important for forming $X(3872)$ state~\cite{Swanson}, which was
also demonstrated by T. Fern\'{a}ndez-Caram\'{e}s and collaborators~\cite{Fern}.
The mechanism also applied to the study of $H$-dibaryon~\cite{QDCSM5}, in which the single
channel $\Lambda\Lambda$ is unbound, but when coupled to the channels $N\Xi$ and $\Sigma\Sigma$,
it becomes a bound state.

\begin{figure}
\begin{center}
\resizebox{0.48\textwidth}{!}{\includegraphics{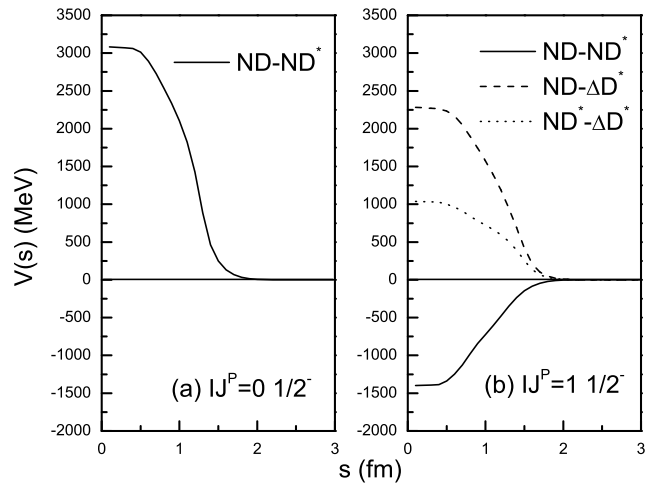}}

\caption{The transition potentials of different channels for the
$ND$ system.}

\end{center}
\end{figure}

\begin{table*}[ht]
\begin{center}
\caption{The binding energies ($E_B$) and the masses ($M$) (in MeV) of every single channels
and those of channel coupling ($c.c.$) for the $NB$ system.}
\begin{tabular}{lccccccc}
\hline
  & & ~~~$NB$~~  & ~~$NB^{*}$~~ & ~~$\Delta B$~~ & ~~$\Delta B^{*}$~~ & ~$c.c.$~\\
\hline
 $IJ^{P}=0\frac{1}{2}^{-}$ & ~~$E_B/M$~~ & $ub/6218$ & $ub/6264$  & $-/-$ & $-/-$ & $-3.2/6214.8$  \\ \noalign{\smallskip}
 $IJ^{P}=0\frac{3}{2}^{-}$ & ~~$E_B/M$~~ &$-/-$ & $-3.4/6260.6$  & $-/-$ & $-/-$ & $-3.4/6260.6$    \\ \noalign{\smallskip}
 $IJ^{P}=1\frac{1}{2}^{-}$ & ~~$E_B/M$~~ &$ub/6218$ & $ub/6264$  & $-/-$ & $-14.5/6542.5$ & $ub/6218$    \\ \noalign{\smallskip}
 $IJ^{P}=1\frac{3}{2}^{-}$ & ~~$E_B/M$~~ &$-/-$ & $ub/6264$  & $-13.5/6497.5$ & $-8.7/6548.3$ & $ub/6264$    \\ \noalign{\smallskip}
 $IJ^{P}=1\frac{5}{2}^{-}$ & ~~$E_B/M$~~ &$-/-$ & $-/-$  & $-/-$ & $-26.1/6530.9$ & $-26.1/6530.9$    \\ \noalign{\smallskip}
\hline 
\end{tabular}
\label{bound_b}
\end{center}
\end{table*}

For the $I=0,~J^{P}=\frac{3}{2}^{-}$ system, it includes only one channel $ND^{*}$, and
it is a bound state with the mass of $2940.3$ MeV, which is close to the mass of
$\Lambda_{c}(2940)^{+}$. Therefore, in our quark model calculation, $\Lambda_{c}(2940)^{+}$
can be explained as a $ND^{*}$ molecular state with the quantum number $J^{P}=\frac{3}{2}^{-}$.
This result is consistent with the conclusion of Ref.\cite{HeJ}, in which they proposed that
the $\Lambda_{c}(2940)^{+}$ could be explained as isoscalar $S-$wave or $P-$wave $D^{*}N$
systems with $J^{P} = \frac{3}{2}^{-}$ or $J^{P} = \frac{1}{2}^{+}$ in the framework of
the one boson exchange model. Meanwhile, a constituent quark model calculation also supported
the existence of $\Lambda_{c}(2940)^{+}$ as a molecular state composed by nucleon and $D^{*}$
mesons with $J^{P} = \frac{3}{2}^{-}$~\cite{Ortega}.

For the $I=1,~J^{P}=\frac{1}{2}^{-}$ system, the $ND$ is unbound because of the repulsive
interaction between $N$ and $D$ as shown in Fig. 1(a). And for $ND^{*}$ channel, the
attraction is too weak to tie the two particles together, so it is also unbound. Due to the
stronger attraction, the energy of $\Delta D^{*}$ is below its threshold, so the standalone
$\Delta D^{*}$ is a bound state here. Then, we do a channel-coupling calculation.
The results show that no stable state can be obtained, i.e., all the energies obtained are
higher than the threshold of $ND$, which indicates that the channel-coupling effect is not
strong enough to make $ND$ bound here. The transition potential of these three channels are
shown in Fig. 3(b), and we find they are smaller than the one of $I=0,~J^{P}=\frac{1}{2}^{-}$
$ND$ and $ND^{*}$ channels, which is shown in Fig. 3(a). Moreover, coupling to the $ND$
and $ND^{*}$ channels, the energy of state $\Delta D^{*}$ is pushed above its threshold,
thus preventing a resonance from materializing.

For the $I=1,~J^{P}=\frac{3}{2}^{-}$ system, the results are similar with those of the
$I=1,~J^{P}=\frac{1}{2}^{-}$ system. The $ND^{*}$ is unbound due to the repulsive potential
between $N$ and $D^{*}$ as shown in Fig. 2(b). Both the standalone $\Delta D$ and $\Delta D^{*}$
states are bound because of the strong attractions between the corresponding two hadrons.
However, these two states disappear by coupling to the $ND^{*}$ channel.

For the $I=1,~J^{P}=\frac{5}{2}^{-}$ system, there is only one channel $\Delta D^{*}$, its
energy, $3210.1$ MeV, is below the corresponding threshold. It is a good resonance state
after coupling to $ND$ by means of tensor interaction. This result is consistent with the one of
Ref~\cite{Cara}, in which they showed that the $\Delta D^{*}$ with $I=1,~J^{P}=\frac{5}{2}^{-}$
was an attractive state, presenting a resonance close to threshold by means of a chiral
constituent quark model.

Because of the heavy flavor symmetry, we also extend the study to the bottom case of $NB$ system,
the numerical results for which are listed in Table \ref{bound_b}. The results are similar to the
$ND$ system. From Table \ref{bound_b}, we find there are several interesting states:
a $NB$ bound state with $I=0,~J^{P}=\frac{1}{2}^{-}$ by two channels ($NB$ and $NB^{*}$) coupling;
a $NB^{*}$ resonance state with $I=0,~J^{P}=\frac{3}{2}^{-}$; and a $\Delta B^{*}$ resonance state with $I=1,~J^{P}=\frac{5}{2}^{-}$.

\section{Summary}

In summary, we perform a dynamical calculation of the $ND$ systems with $I=0$ and $1$,
$J^{P}=\frac{1}{2}^{\pm}$, $\frac{3}{2}^{\pm}$, and $\frac{5}{2}^{\pm}$ in the framework
of QDCSM. Our results show:
(1) All the positive parity states are unbound in our calculation.
(2) The pure $ND$ with $I=0,~J^{P}=\frac{1}{2}^{-}$ is unbound, but a bound state with
mass of $2801.0$ MeV can be obtained by coupling the $ND^{*}$ channel. The mass of this bound state
is close to the observed $\Sigma(2800)$, which shows that $\Sigma(2800)$ can be explained as a $ND$
molecular state with the quantum number $J^{P}=\frac{1}{2}^{-}$ in our quark model calculation.
(3) The $ND^{*}$ with $I=0,~J^{P}=\frac{3}{2}^{-}$ is also a resonance state with the mass of
$2940.3$ MeV, closing to the mass of $\Lambda_{c}(2940)^{+}$, which indicates that
$\Lambda_{c}(2940)^{+}$ can be explained as a $ND^{*}$ molecular state with the quantum number
$J^{P}=\frac{3}{2}^{-}$ in QDCSM.
(4) The $I=1,~J^{P}=\frac{5}{2}^{-}$ $\Delta D^{*}$ is also a resonance state with mass
of $3210.1$ MeV.
Besides, the calculation is also extended to the bottom case of $NB$ system. The results
are similar to the case of the $ND$ system. On the experimental
side, confirming the existence of the charmed hadrons $\Sigma(2800)$ and $\Lambda_{c}(2940)^{+}$
is an interesting subject. Besides, searching for other molecular states with heavy quarks,
such as $\Delta D^{*}$, $NB$, $NB^{*}$ and $\Delta B^{*}$ will be challenging topics in future.

\section*{Acknowledgment}
This work is supported partly by the National Science Foundation
of China under Contract Nos. 11205091, 11035006 and 11535005, the Natural Science Foundation of
the Jiangsu Higher Education Institutions of China (Grant No. 16KJB140006), and Jiangsu Government
Scholarship for Overseas Studies.

\end{document}